\documentclass{article}
\usepackage{graphicx, caption} 
\usepackage[a4paper, total={6in, 8in}]{geometry}
\usepackage{url}
\usepackage{authblk}
\usepackage{tikz}
\usepackage[numbers]{natbib}

\newcommand{\Winterlab}{\textit{WinterLab} }

\newcommand\copyrighttext{%
  \footnotesize Copyright © 2023 by the International Journal of Designs for Learning, a publication of the Association of Educational Communications and Technology. (AECT). Permission to make digital or hard copies of portions of this work for personal or classroom use is granted without fee provided that the copies are not made or distributed for profit or commercial advantage and that copies bear this notice and the full citation on the first page in print or the first screen in digital media. Copyrights for components of this work owned by others than IJDL or AECT must be honored. Abstracting with credit is permitted.}
\newcommand\copyrightnotice{%
\begin{tikzpicture}[remember picture,overlay]
\node[anchor=south,yshift=10pt] at (current page.south) {\fbox{\parbox{\dimexpr\textwidth-\fboxsep-\fboxrule\relax}{\copyrighttext}}};
\end{tikzpicture}%
}

\title{\textit{WinterLab}: Developing a low-cost, portable experiment platform to encourage engagement in the electronics lab}

\author{Maclean Rouble \and Matt Dobbs \and Adam Gilbert}

\date{March 2023}

\begin{document}

\maketitle

\begin{abstract}
    Encouraging student engagement is a key aim in any educational setting, and allowing students the freedom to pursue their own methods of solving problems through independent experimentation has been shown to markedly improve this. In many contexts, however, allowing students this flexibility in their learning is hampered by constraints of the material itself, such as in the electronics laboratory, where expensive and bulky equipment confines the learning environment to the laboratory room. Finding ourselves in the position of teaching one such laboratory course at the undergraduate level, we sought to encourage students to learn through independent investigation and the pursuit of personal projects, by providing a more flexible and inquiry-based learning environment and allowing them to take their measurement equipment – and their learning – beyond the laboratory itself. We present this project as a case of design both for and by students, with the lead designer undertaking the project after attending the course in question, and pursuing its development as a foundational step in their graduate career. We discuss the challenges and opportunities we encountered over the course of the design and development process, and the eventual key output of the project: a portable, low-cost, integrated electronics experimentation platform called the \Winterlab board. 
\end{abstract}

\copyrightnotice

\paragraph*{Maclean Rouble} is a PhD candidate in the Department of Physics at McGill University. Her research focuses on the development of instrumentation and electronic readout architectures for millimeter-wavelength telescopes.

\paragraph*{Matt Dobbs} is a Canada Research Chair in the Department of Physics and associate member of the Department of Electrical and Computer Engineering at McGill University and is a senior fellow in the Canadian Institute for Advanced Research Gravity \& the Extreme Universe program.

\paragraph*{Adam Gilbert} (at the time of development of this project) was an academic associate working as an electrical engineer in the Department of Physics at McGill University. He now works at Nüvü Caméras Inc.

\section{Introduction}

The McGill Cosmology Instrumentation Laboratory, often casually referred to as the Winterland Cosmology Lab, is a research group at McGill University specializing in the design, development, and operation of telescope systems at radio and microwave frequencies. Headed by Prof. Matt Dobbs, the group is comprised of professional scientists and engineers, post-doctoral researchers, and students at both the graduate and undergraduate levels. Members typically undertake projects related to electronic design or statistical analysis of telescope data. In this case, however, we present one of the group's projects that, while grounded in electronic design, took its motivation and destination from a rather different source: encouraging junior students of electronics to learn through tinkering and independent experimentation, by providing them with a custom-built portable electronics learning platform that achieves the functionality of the test and measurement equipment typically found in an undergraduate electronics laboratory.

The project was first suggested by Prof. Dobbs to then-undergraduate student Rouble, who had expressed interest in joining the research group. Inspired by a commercially available solderless prototyping breadboard with an integrated Arduino-type microcontroller, the idea was to turn such a product into a portable platform that could replace the test and measurement equipment typically used in the laboratory component of an electronics and signal processing course taught by Dobbs to second-year undergraduates. Feeling that true internalization of learning in electronics (as it does in many domains) comes through independent experimentation, but finding that difficult to encourage within the traditional recipe-based laboratory activities of the course, he proposed providing such an experiment platform to students. This, perhaps accompanied by a more open-ended curriculum, would free them from the confines of the laboratory space and encourage them to undertake their own personal projects. Rouble's pursuit of this project, for a course that she had herself taken the previous year, would serve not only to examine the potential of this platform as an avenue for curriculum and learning environment development but also as an initiation to scientific work with the rigor expected of members of the research group. Although at its proposition, the scope of the project was limited to a short technological demonstration, the team’s engagement with the project and belief in its potential value to other students fostered an enthusiasm to pursue it beyond its initial goals. Undertaking it as a mid-level undergraduate, Rouble chose to continue further development of the concept and its application for several years, taking it beyond the completion of her degree and into the next, and putting to good use the skills and confidence acquired in its pursuit.

We describe the full design process of this project, not only to confer the techniques needed to build the electronics learning platform that it eventually produced, but also to share the challenges, abandoned avenues, and successes we encountered along the way. This is as much a story of personal and academic growth as it is of technological design: of a junior scientist, equipped with basic principles and ideas, building and expanding on them in order to design, improve upon, and finally create a refined, operational device intended to encourage students in similar contexts to venture forth in the same ways, to reach their own full potentials.

\section{Motivation}
\subsection{The Pitch}
Engaging with research early in an academic career can have many advantages for students. Not only does the experience reflect well when listed on a resume, but the skills acquired are often distinct from those taught in the classroom, and the interactions with more experienced researchers help to shape the student’s trajectory in the field at large. On the side of the research group, a challenge may lie in finding a project which balances suitability to the student in terms of duration and scope, with making a genuinely useful contribution to the group’s body of work. The case described in this article was initially conceived as one such student project.

When Rouble, then finishing the second-to-last year of an undergraduate degree, expressed interest in doing work with the Winterland Cosmology Lab, Dobbs proposed a trial project which could span perhaps a semester or two in length, giving Rouble a taste of work with the group.

Based on his experiences and that of colleagues, he surmised that truly gaining familiarity with electronics and circuit building – key elements of work in the group – came from undertaking personal projects and having fun tinkering, but within an introductory circuits course he regularly instructed, he found it difficult to provide students with opportunities for this type of engagement. The course featured a laboratory component in which students would learn to construct basic circuits, use test and measurement devices such as solderless breadboards, signal generators, oscilloscopes, and multimeters (see Figure \ref{fig:fig1}), and perform experiments to characterize simple circuitry made of unknown components.

While many of the tools used in this electronics laboratory were useful for hobby projects, they were confined to the laboratory itself and few, if any, students made use of them for extracurricular projects. Having these tools at a student’s disposal at home might encourage them to incorporate them into hobby projects. To enable this, however, would require a replacement for the traditional laboratory equipment, one which:
\begin{itemize}
    \item Provided effective, easy-to-use implementations of the key test and measurement tools used in the lab.
    \item Was portable and robust.
    \item Was inexpensive, so that each student could have and keep one beyond the end of the course.
\end{itemize}

He had taken note of new educational tools such as the STEMTera$^{TM}$ board (see Figure \ref{fig:fig2}), a commercially-available prototyping board with an integrated Arduino-style microcontroller and wondered whether it could be purposed for this use. It would take some effort to develop it and determine its viability, and with a background in hobbyist electronic design and microcontroller programming, Rouble seemed a good fit for the project. She enthusiastically agreed to take it on.

Although conceived as part audition piece and part technology demo, the scope of the project quickly grew into something more profound. As the design team, Rouble, Dobbs, and Gilbert experimented with the technological side of the project and began to appreciate its strengths and limitations, we also delved into teaching and education literature to educate ourselves, as none of the research group members involved in the project have backgrounds in learning theory. We used this study of the literature to motivate and better understand how we could put into practice what we intuited to be true: that learning circuits and signal processing was best done through tinkering and independent experimentation, and how we could use the project to offer this opportunity to students in the lab. In the next section, we provide an overview of the theoretical background and research that we found most compelling, and which most helped to shape the design of this project.

\subsection{Theoretical Framework}
\begin{figure}
    \centering
    \captionsetup{width=0.8\linewidth}
    \includegraphics[width=0.7\textwidth]{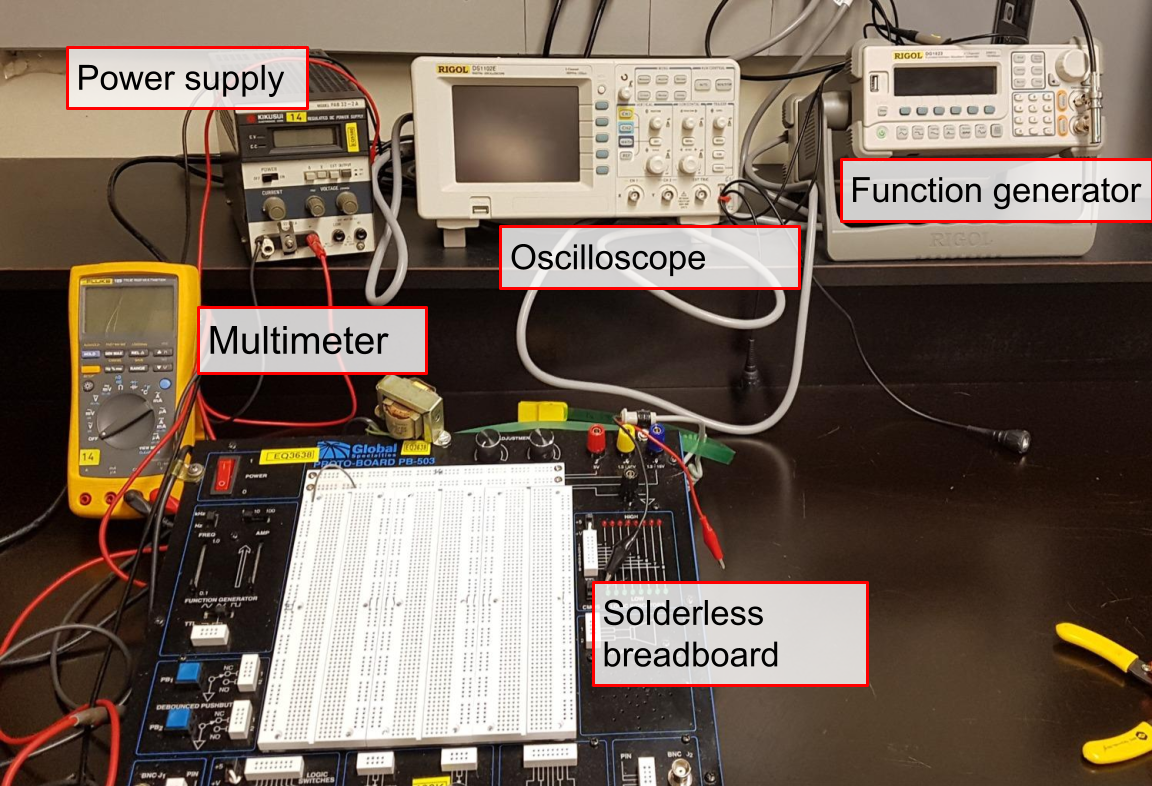}
    \caption{A test and measurement station in the laboratory used in the undergraduate signal processing course discussed in this project. Students have access to a wealth of equipment (multimeter, power supply, oscilloscope, function generator, and solderless breadboard) but can only interact with it within the laboratory itself during limited opening hours.}
    \label{fig:fig1}
\end{figure}

Our motivating intuition that learning is best done through independent and practical experimentation is a well-established principle in education literature and is not limited to the fields of electronic circuitry or signal processing. \citet{appleton2008} find that maximizing student engagement with the subject matter of a course is a key factor in improving student outcomes, and it has been repeatedly shown that the traditional teaching-by-telling style (\citet{freeman2014}) is not as effective in this regard (\citet{bransford1999}; \citet{mcdermott1993}). Rather, students who are afforded opportunities to interact with new knowledge in self-ascribed ways strongly outperform their traditionally-instructed peers (\citet{freeman2014}; \citet{michael2006}; \citet{thornton1998}). Despite this, many courses still follow the traditional style of teaching, and this disconnect is particularly apparent in laboratory courses. Science laboratory courses are often entrenched in traditional styles, in part due to the high cost of the laboratory equipment that is typically located in a limited-access teaching laboratory, a trend from which the introductory signal processing course in question in this project was no exception – an example student laboratory setup is shown in Figure \ref{fig:fig1}.

In recent years, low-cost, commercially available components have begun to offer options addressing this apparent disconnect in the form of computer or mobile phone applications and microcontroller-based gadgets, which enclose multi-purpose measurement equipment in a small, affordable, easy-to-use package (e.g., the iOLab). These tools are intuitive for students to use, and so allow more time to be spent learning the course subject matter. Many are also designed to make laboratory activities more fun, and to allow students to work where they please: these devices are highly portable, turning what was once a prescribed sit-down session in an often dark, basement laboratory into an investigation that can be performed anywhere the student sees fit. 

This technology-enabled learning enrichment can be taken further, by creating laboratory activities in which students create not only their own strategies but also their own measurement equipment, demonstrations and documentations of which may be found freely across the internet as well as in academic literature (e.g., \citet{bouquet2017}). The key ingredient in many of these low-cost solutions is the microcontroller, a small computer typically mounted as part of a circuit board with multiple in- and outputs through which it interfaces with external circuitry and measuring devices. Popular among hobbyists as well as in the learning laboratory, they provide a useful foundation for individual experimental setups, such as those presented by \citet{zachariadou2012} and \citet{schultz2016}, among others. The accessibility of these devices and their easy integration into experiments is a departure from the black-box nature of traditional measurement equipment, as their workings and those of the test setups they enable may be more readily understood by students at most levels.

Drawing on inspiration from the above and other examples, it was our aim to develop and provide a low-cost, portable, integrated experiment platform that students could own and continue to use after the end of the course. We hoped – and sought to evaluate whether – this would support a fun, engaging learning environment and foster enthusiasm and self-motivated skill-building in electronics, abilities which may be carried over into the sciences at large.

\section{Early Endeavors}
It will likely not surprise the reader that an undergraduate striving to implement the functionality of a full electronics laboratory test and measurement setup within a pocketbook-sized computing device would not be fully successful on a first attempt. The initial prototype, though it never achieved sufficient performance to be used in the course for which it was intended, laid the groundwork for future iterations of the device. Many aspects of its design were directly ported into the later version, and the course of their initial development in this first prototype allowed us to flesh out and explore the nature of the device we had decided to create. We tested the prototype by using it to complete the basic aspects of a typical course laboratory. This allowed us to update the requirements to include practical aspects that were not apparent at project initialization.

\begin{figure}
    \centering
    \captionsetup{width=0.7\linewidth}
    \includegraphics[width=0.5\textwidth]{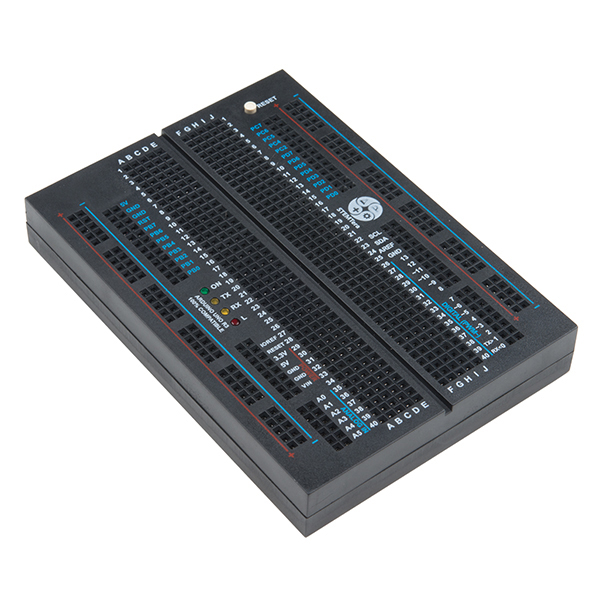}
    \caption{The STEMTera$^{TM}$ board (image licensed under CC BY 2.0). An Arduino-style microcontroller, the ATmega328P, is embedded within the breadboard, allowing the user to prototype projects with convenient access to the programmable inputs and outputs.}
    \label{fig:fig2}
\end{figure}

For the initial prototype, as suggested in Dobbs’ project pitch, we began by simply re-programming the commercially-available STEMTera$^{TM}$ (see Figure \ref{fig:fig2}) microcontroller breadboard, imbuing it with the ability to synthesize and digitize waveforms, to supply voltages, and to measure resistance, voltage, and capacitance in a circuit under test. Using this off-the-shelf device had the appeal of not requiring any manufacturing on our part, presenting a physically self-contained and attractively-styled experiment platform. The microcontroller at the heart of the device, the ATmega328P (the same device that powers the Arduino Uno \cite{arduino_uno}), would allow control of the device via USB from a student's laptop, while the integrated solderless breadboard would allow students to build their experiment circuits directly onto it: altogether a convenient, compact experiment package. To accompany the board, we began work on a software graphical user interface, which would display the signals measured by the device and allow the user to control the board’s functions.

Rouble led the work on these aspects of the project, under the supervision of Dobbs and frequently sought electrical engineering guidance from Gilbert. Although familiar with simple circuit design and microcontroller programming, and having pursued courses in computer science, both aspects of the project presented challenges to the student, requiring significant time spent studying technical memoranda on a range of topics from digital signal processing to user interface coding frameworks – skills which have continued to be relevant in subsequent work.

With these developing skills in hand, the project progressed, but it became apparent to us that using this product would present some critical difficulties. The low internal clock rate of the ATmega328P severely limited its ability to process data and synthesize and digitize analog signals. Because the device's main functionalities would all be performed internal to the microcontroller, this meant that these clock cycles were divided among the simultaneous processes of synthesizing and digitizing voltage waveforms (both of which require sampling at a rate in at least the thousands of times per second), transmitting the digitized data over a serial connection to the user's laptop, and regularly receiving user commands coming from the user's laptop over the same serial interface. Waveform synthesis was somewhat expedited by the addition of an external parallel digital-to-analog converter, but to do even the most central three tasks in sequence (output a voltage value at its waveform generator pin, measure a voltage at its oscilloscope input pin, and transmit that single measured value to the user’s laptop) took nearly 0.4ms to execute, and would provide just one datapoint to be displayed to the user. This limited the sampling rate of the function generator and oscilloscope to less than 3kHz: much too slow for the typical circuits studied in undergraduate laboratories. 

This led us to redesign the program execution loop: rather than run each task in a continuous sequence, instead, it would run the synthesis and digitization tasks for some period (say, half a second), during which time a waveform is outputted from the function generator and a digital buffer is populated with measured values from the oscilloscope. When the buffer is full, we pause data acquisition and transmit the entire buffer at once to the user's computer. Although this allowed us to generate waveforms at an order of magnitude higher frequency, during the transfer of the oscilloscope’s buffer to the computer, the board’s other functions would pause, leading to a waveform that blinked on and off.

These limitations on the range of frequencies accessible by the device led us to formalize our goals for this metric: if we could produce a device capable of synthesizing and digitizing waveforms up to 22kHz, we would span the audible spectrum, enabling students with a musical bent to use the device for sound engineering projects. Although we had not previously formalized the frequency range that would satisfy our goals, meeting, or exceeding, this number became our design target, and choices of technology for the next stages of the project were made with this in mind. This would also allow us to design laboratory modules in the audible range, where students could both see (using the oscilloscope) and hear (using a speaker and their ears) the circuit’s output. 

\begin{figure}
    \centering
    \captionsetup{width=0.7\linewidth}
    \includegraphics[width=0.6\textwidth]{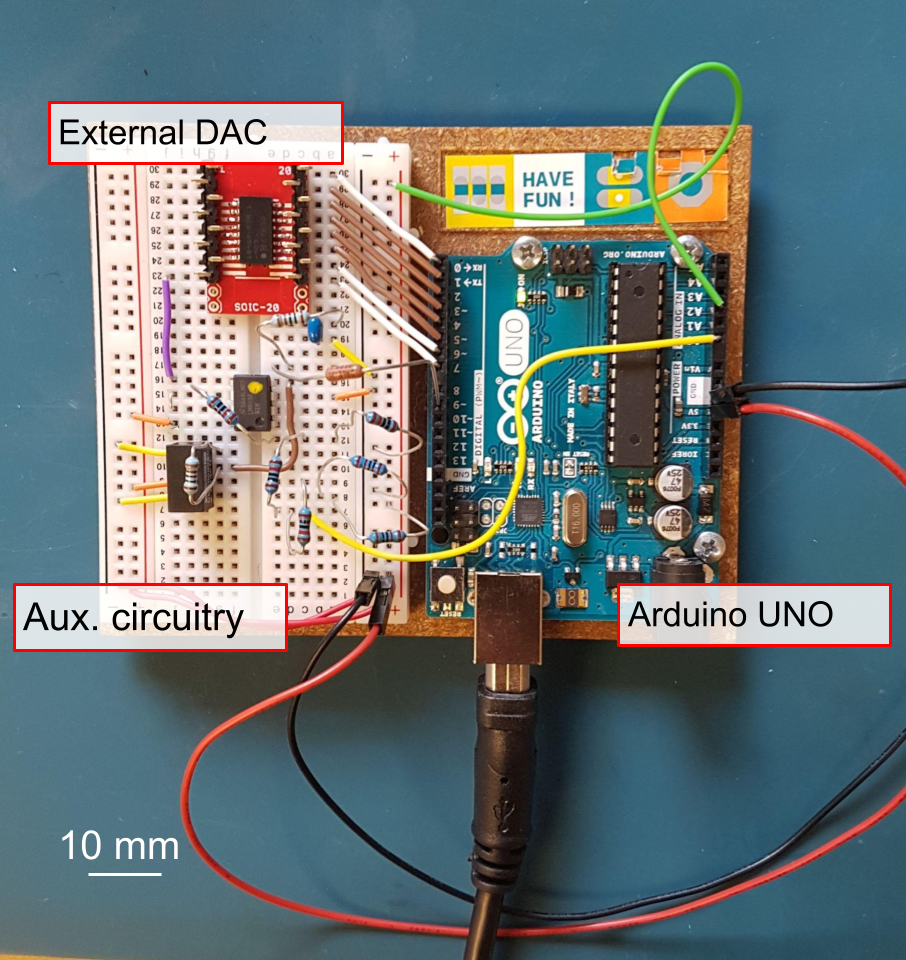}
    \caption{The initial prototype of the custom electronics experimentation platform, built using an Arduino UNO. The external auxiliary circuitry, including the external digital-to-analog converter (DAC), required to synthesize waveforms and protect the microcontroller from accidental over-voltages led us away from our initial plan to use the streamlined STEMTera$^{TM}$. }
    \label{fig:fig3}
\end{figure}

By this point, however, we had had to add an external piece of circuitry to the STEMTera$^{TM}$ board: the digital-to-analog converter. This was already a strike against our initial plan to have the device as self-contained and streamlined as the product was when it arrived in its mail-order box. We recognized that, in order for students to be able to measure a standard range of voltages in the laboratory, such as -5 to +5V, we would need additional external circuitry to scale the input and output voltage to suit the smaller range that could be handled by the microcontroller itself. Further, we recognized that, in order to allow students to experiment with confidence, we would have to ensure that the eventual experiment platform was robust enough to survive an occasional accidental large voltage being applied to its ports. This meant the addition of even more external circuitry, in the form of buffer amplifiers to stand between each of the microcontroller's inputs and outputs and the device's connections to the outside world. This rapidly growing network of external circuitry (seen on the breadboard in Figure \ref{fig:fig3}), combined with the heavy limitation of the device's sampling rate, was the final strike again our initial vision of the experiment platform. While more experienced eyes might have noticed these limitations from the outset, discovering them firsthand was a valuable lesson for Rouble, one that would be applied throughout the rest of the design process.

\section{Refinement}
Having decided to shift the hardware basis for the platform away from using a pre-assembled board such as the STEMTera$^{TM}$, we were afforded more freedom in our specifications, informed by several months spent developing the first prototype and supported by the research group’s expertise in electronic systems and circuit design.

First, we needed to select the microcontroller around which we would design the platform. In working with the ATmega328P, we had recognized the importance of using a chip that could be programmed using the Arduino language and software. These tools are designed to be user-friendly and are widely documented on the internet. We had coded the firmware for the prototype in straightforward terms so that a user with a facility with programming typical of an undergraduate could modify it, and felt that this extensibility in the platform's potential as a learning tool was necessary to preserve in the final design. Fortunately, the Arduino language and programming platform are widespread, and we chose a chip that not only boasted a much faster (180MHz) onboard clock but also was associated with a vibrant online support forum populated by enthusiasts and professionals alike – a motivating factor in our choice – which was helpful during our development of the project and might assist future users in their endeavors. Called the Teensy$^{\textcircled{R}}$ 3.6, this device is a compact development board that supports a Kinetis K-66 family microcontroller\cite{nxp}, offering an array of input and output pins as well as a USB interface to connect to a computer\cite{stoffregen}. It also offered the capability to utilize a direct memory access (DMA) buffer, allowing storage of data without the use of interrupts within the main execution loop. Having seen the utility of speeding up the prototype's operation with the use of buffered waveform samples, we recognized that these DMA channels could be used to output waveform samples at high speed without monopolizing clock cycles to the detriment of the rest of the processor's operation. 

\begin{figure}
    \centering
    \captionsetup{width=0.8\linewidth}
    \includegraphics[width=0.7\textwidth]{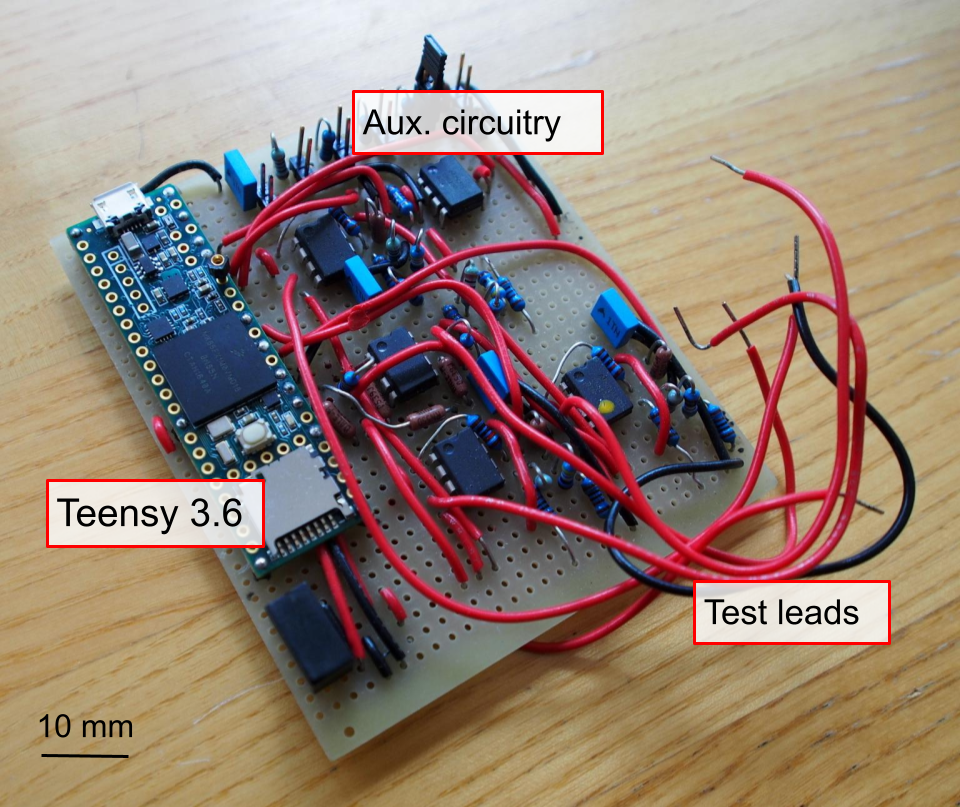}
    \caption{A hand-soldered breadboard prototype of the \Winterlab board. Identical in functionality to the final version, this prototype served as a useful testbed and was used by a volunteer group of students in the 2018 edition of the signal processing course to test and debug the concept.}
    \label{fig:fig4}
\end{figure}

Some aspects of the initial prototype could be directly incorporated into the new design, such as the external voltage scaling and protection circuitry, and with the new processor's higher sample rate, we no longer needed the external DAC. We soon began building our first evaluative versions of the device. After sketching out our circuit designs on paper and in simulation, we transferred them to a solderless breadboard. This allowed us to verify our choice of components, and also to develop an idea of how best to lay out all the elements for the next phase of prototyping, the hand-soldering of a fully functional version of what we had begun to refer to as the \Winterlab board (see Figure \ref{fig:fig4}).

Two such handmade prototypes were constructed and given to a small group of student volunteers from the 2018 edition of the introductory signal processing course for which the \Winterlab was designed. They used the board to complete a subset of the course's laboratory activities, with the designers (primarily Rouble) present to discuss its performance, debug issues, and hear their feedback as they worked. 

We encountered several issues during these sessions, in both the hardware of the prototype and the software that accompanied it. For example, the test leads, which on the prototype were simple wires soldered directly to the board, were too short and stiff, making it awkward to make connections with the circuits under test in the laboratory activities. To address this, we procured a new set of leads that were longer, made of flexible cable, and had spring-loaded grabber hooks at each end. These were quick and easy to attach to measurement setups during laboratory activities. We incorporated this style of test lead into our final design of the platform, changing our input/output connection points into metal loops onto which the grabber hooks could be easily attached.

Although we had tested the initial software accompanying the prototype platform by first completing the laboratory activities ourselves, we had not tried to install it on a wide range of computers, and quickly ran into bugs and error messages when attempting to get it working for the trial sessions. This made it clear that any software that we distributed to a larger group of users would need to be completely self-contained and compatible with all common operating systems, to avoid students struggling to install it on their own laptops. 

Even after getting it running during the trial sessions, the volunteers found the initial draft of the software interface to be unresponsive, unaesthetic, slow, and without much resemblance to physical laboratory measurement equipment. Lagging in the interface caused frustration and impeded their ability to easily use the platform to record data. Through reorganization and streamlining of the interface code, we were able to greatly reduce this lag in the early weeks of the trial sessions, which improved the user experience for the students as the trial continued. We took the students’ feedback on the visuals and interactive experience of the interface into consideration as we worked on the final version of the software: we replaced sliders with clickable dials, enabled entry boxes, and added up/down buttons for quick adjustment of voltage values. In redesigning the interface, we also changed to a different graphical software framework (from Tkinter to pyQt5), to improve its overall appearance.

Despite the challenges encountered while working with the prototype, in an anonymous survey offered to them after the trial sessions to express their opinions on the experience, the student volunteers praised the all-in-one nature of the device and its ability to directly interface with their computers and indicated that if such a device were provided along with the course, they would be likely to use it for projects outside the curriculum.

Informed by this live testing with members of our intended audience, we felt ready to incorporate the lessons we had learned as we moved into the final stages of production of the experiment platform: its design and manufacturing as a printed circuit board (PCB). The layout of the PCB was done by co-author Gilbert, who also arranged its production and population with our chosen components by an outside circuit board manufacturing company. After procuring the remaining hardware that would accompany the board, including a set of grabber-hook test lead cables, a solderless breadboard, and a USB cable, we now had a small fleet of \textit{Winterlab}s, bundled into self-contained kits and ready to be deployed in the undergraduate electronics laboratory.

\section{The \Winterlab platform}
Coming together more than a year after the project was initially proposed, the final version of the \Winterlab electronics learning platform consists of a microcontroller-based development board, built to interface with a solderless prototyping breadboard, and a cross-platform graphical software interface. It provides the user with implementations of the instruments found in a typical science undergraduate-level electronics laboratory: oscilloscope, function generator, digital multimeter, and voltage source. Putting the lessons learned from earlier prototype versions into play in its design, the final product is a robust, compact, low-cost, and highly versatile learning device, for which we have made all electronic and circuit board design files, as well as source code for the microcontroller and user interface, open-source and freely available online.

\begin{figure}
    \centering
    \captionsetup{width=0.8\linewidth}
    \includegraphics[width=0.7\textwidth]{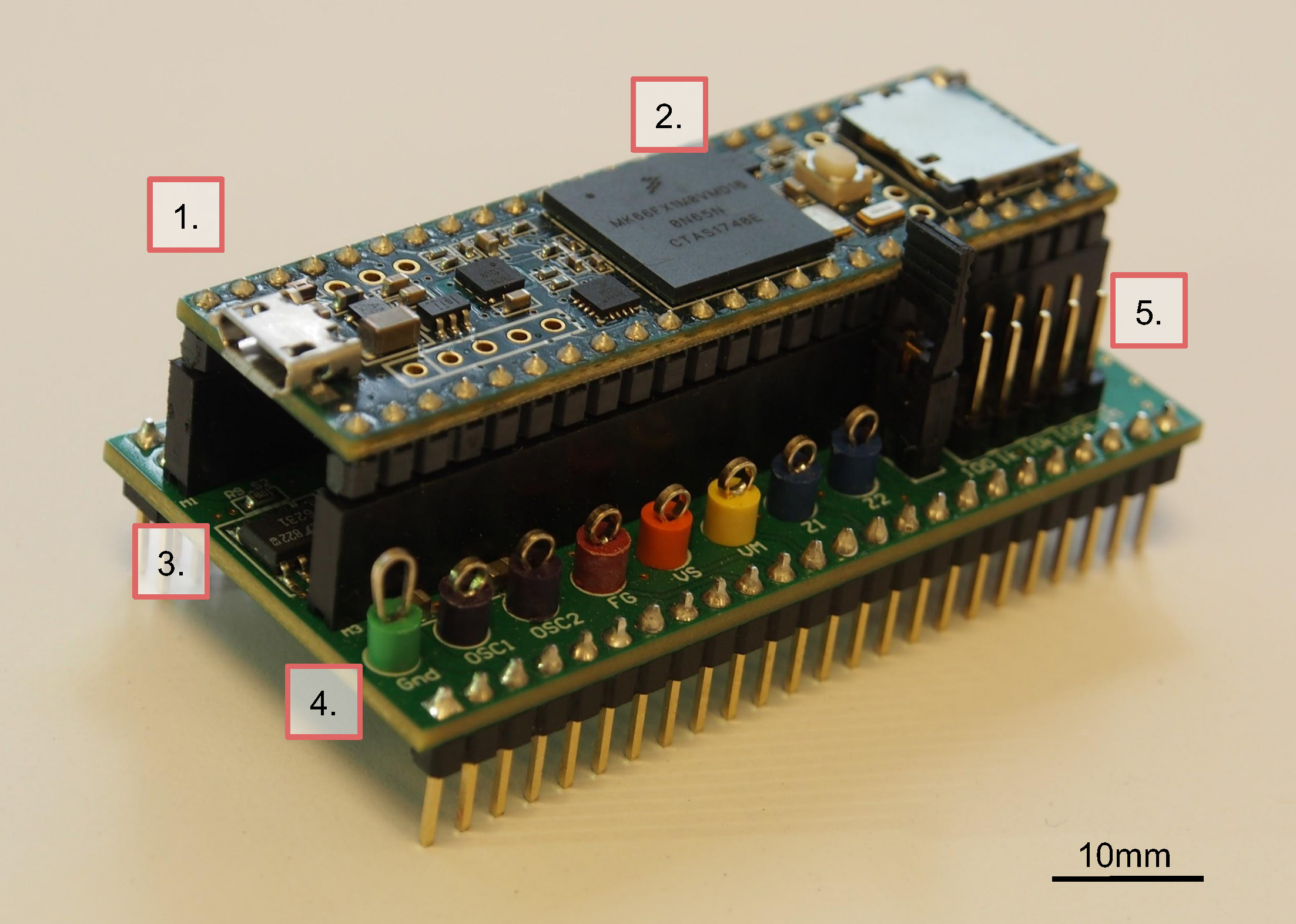}
    \caption{A manufactured \Winterlab electronics experimentation board. The top deck consists of an off-the-shelf Teensy$^{\textcircled{R}}$ 3.6 microcontroller (1), which performs the onboard processing inside the Kinetis K66-family chip (2). The custom-made lower deck (3) contains auxiliary circuitry as illustrated in Appendix A. Connections to and from the board are made by attaching test leads to the row of eight colorful input/output connection points along the lower deck (4). The row of pin pairs at the right of the lower deck (5) serves as a manual measurement scale selector for impedance measurements.}
    \label{fig:fig5}
\end{figure}

\subsection{Hardware}
The hardware element of the \Winterlab platform is the \Winterlab board (see Figure \ref{fig:fig5}). It is comprised of our custom-designed PCB and the Teensy$^{\textcircled{R}}$ 3.6 microcontroller development board, which can be removed from the \Winterlab to be transplanted onto other projects. It has eight input/output connection points, whose functions (depicted in the functional diagram in Appendix A) are, from left to right in Figure \ref{fig:fig5}:
\begin{itemize}
    \item A ground reference.
    \item A two-channel, 8-bit, 240 kSPS oscilloscope.
    \item An 8-bit, 4.1 MSPS arbitrary waveform generator.
    \item A direct current (DC) voltage source.
    \item A voltmeter.
    \item An impedance meter for resistance and capacitance (uses two connection points).
\end{itemize}

\begin{figure}
    \centering
    \captionsetup{width=\textwidth}
    \includegraphics[width=\textwidth]{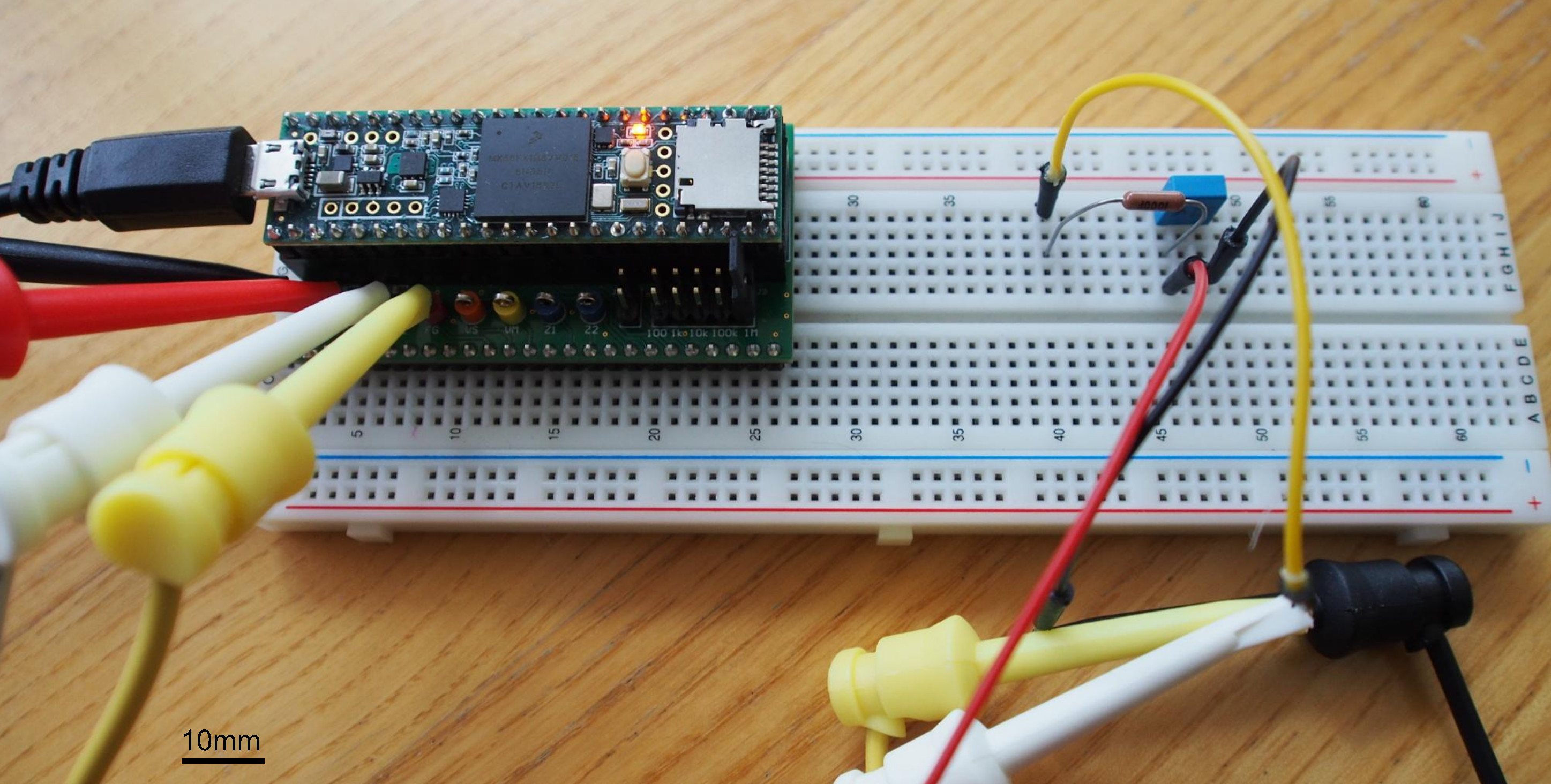}
    \caption{The \Winterlab platform in use. The board is mounted on a solderless breadboard (white), on which a simple filter circuit has been constructed as an example student project (at right). Test leads connect the function generator and two oscilloscope channels to measurement points in the circuit. The board is connected via USB to the user’s laptop, where, on the graphical user interface (see Figure \ref{fig:fig7}), output values can be adjusted, and data recorded.}
    \label{fig:fig6}
\end{figure}

The user connects these inputs or outputs to their system via test-lead cables (see Figure \ref{fig:fig6}). 

The board is designed to be mounted on a solderless prototyping breadboard, a common medium for implementing projects in course-based labs. The user can build their projects directly onto the breadboard, allowing easy connections to and from the \Winterlab, which communicates with and is powered by the user's computer via USB. 

The lower deck of the board, on which the inputs and output connection points are mounted, contains auxiliary circuitry to support the board's function. This is a highly compact, tidy copy of the jungle of wires and through-hole components visible in the hand-soldered prototype of Fig. \ref{fig:fig4}. These classes of circuits are taught at the introductory undergraduate level, require commonly-used parts only, and, although more compact in this machine-manufactured version, can be readily assembled by hand. The cost of the components of the board, including the Teensy$^{\textcircled{R}}$ 3.6 microcontroller, is comparable to a typical undergraduate textbook. We have made the detailed schematics for these circuits available with the source code and other useful system files at \url{github.com/wl-base/\Winterlab}.

\subsection{Software}

\begin{figure}
    \centering
    \includegraphics[width=\textwidth]{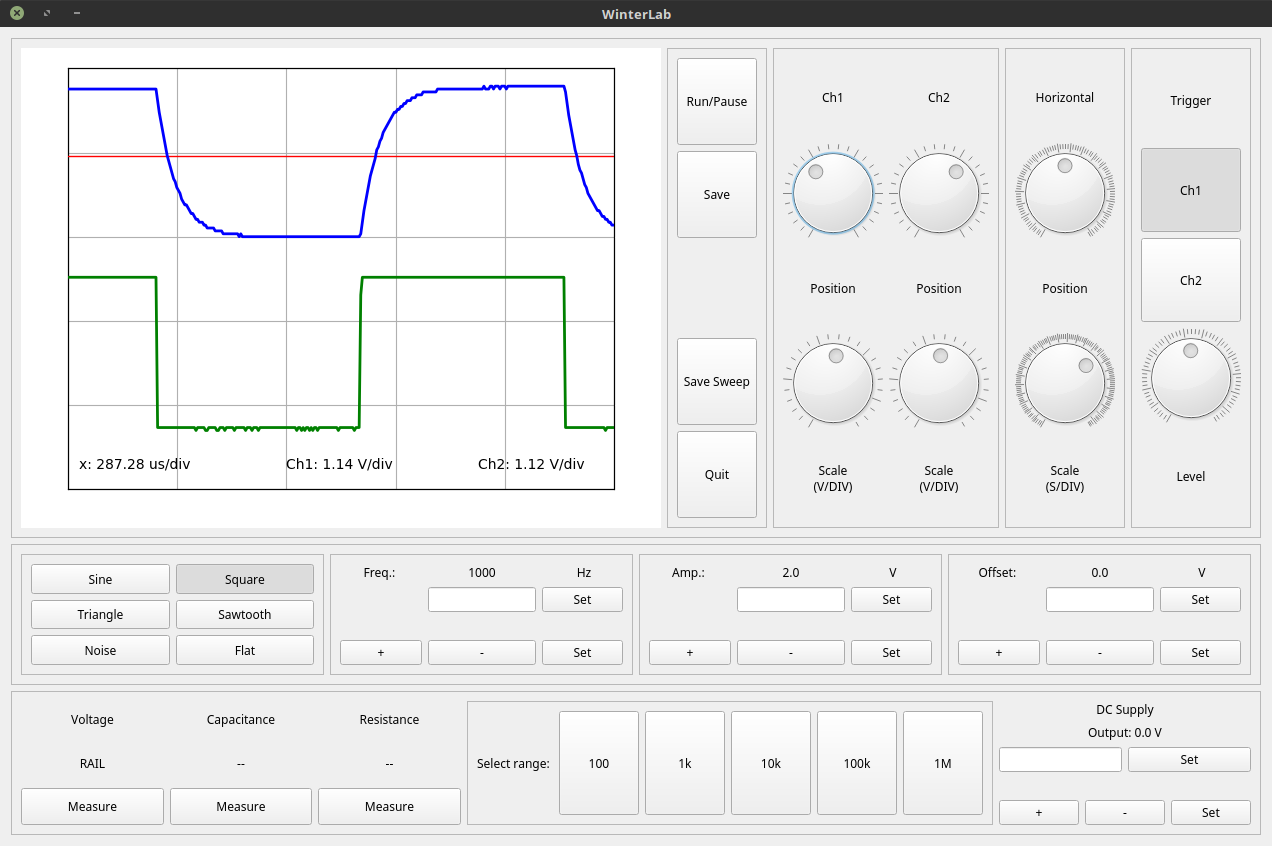}
    \caption{The WinterLab graphical user interface (GUI). This cross-platform program runs on the user’s personal computer and controls the WinterLab board via USB. It consists of three main sections, each corresponding to the measurement device whose functionality it implements: the oscilloscope (top), the waveform generator (middle), and the digital multimeter with DC voltage source (bottom). Here, the display corresponds to the simple filter measurement shown in Figure \ref{fig:fig6}.
}
    \label{fig:fig7}
\end{figure}

The \Winterlab software consists of a graphical user interface (GUI; see Figure \ref{fig:fig7}), which controls the \Winterlab board when connected to the user's computer via a USB cable. The GUI is a cross-platform program tested on Mac OS, Linux Ubuntu, and Windows (including Windows touchscreen tablet PCs), whose source code is written in Python. Although Python is an immensely popular language, especially in the sciences, to meet our goals for portability and ease of use, we considered it important that the user need not know anything about programming in order to use the interface, so we were careful to build the GUI as a self-contained program that can be run at the click of a mouse. For the interested user, the Python-based code is easily understood, and therefore readily modifiable. 

We sought to have the appearance of the interface emulate that of traditional rack-mounted laboratory test and measurement equipment, to reinforce the link between our platform's implementation of these tools and their physical counterparts. Each of the devices implemented within the platform (oscilloscope, function generator, voltage source, digital multimeter) is represented in the display, and clickable buttons, dials, and input fields allow control of the platform's input and output functionalities. The interface also allows automated functions, such as generating and recording a frequency sweep, a common test used to characterize many types of circuits which is time-consuming when done by hand, and which students in the course had frequently expressed a wish to automate. Combining this type of automation with the ability to save acquired images and numerical data directly onto the user's laptop, we aspired to create an interface that is both convenient and easy to use for student lab projects.

Like the hardware, the software interface underwent a major overhaul midway through the design process. At the outset of the project, we chose a GUI code backend that was simple to learn and adapt, with limited knowledge of the other available options, and without carefully designing the requirements for a user-friendly student experience. A functional version was created, and given to the student volunteers to use during the testing stages for the handmade prototype hardware. While polite in their constructive criticism, they found it clunky and unintuitive – insights we took to heart as we moved forward in the design process.

After the semester spent testing the prototypes with the volunteers, we refactorized the GUI with a new, more modern Python backend, pyQt5. Although this switch required considerable research and practice to become familiar with its syntax and structure, it was worth the time invested, as the resulting program is robust and considerably more attractive than the first attempt. Engineering it to run smoothly and without crashes drew on Rouble’s coursework experience, which had recently included topics emphasizing designing robust software. In keeping with the development of the hardware, the pursuit of this aspect of the project afforded her additional chances to engage with skills that had been learned, but not yet applied in practice.

\section{Field Testing}
Motivated by our desire to engage students with their learning and to encourage them to pursue projects on their own, we aspired to also find a way to quantitatively evaluate what impact the changes we made to the environment of the laboratory and the delivery of its curriculum had on the students who participated in it. Our plan for this evaluation was straightforward, and in hindsight, perhaps naively so, chiefly due to a failure to anticipate the difficulty in acquiring volunteers and in entreating them to consistently complete our study materials.

We developed and deployed a new inquiry-based laboratory curriculum, enabled by the \Winterlab platform, in the edition of the introductory signal processing course that ran in the 2019 school year. Without altering the topics explored, the modified curriculum instead presented the lab activities as puzzles with short explanations, rather than as detailed recipes. This delivery style placed more importance on students’ experimentation than on the acquisition of a particular “correct” result. The activities were intended to encourage students to explore various avenues and to collaborate amongst themselves, with the majority of marks awarded for thorough documentation and discussion of the experiment process itself.

The class comprised 113 students and has both lecture and laboratory components. For the laboratory portion, the students are typically subdivided into five sections, each completing the activities in separate sessions. Before students had enrolled in the course, we designated one such section to be the "pilot" session, in which we would deploy the modified curriculum and \Winterlab platform, while the other sections would continue to follow the traditional style of instruction. We planned to monitor the experience of students in both the pilot and traditional sections with the use of a series of anonymized surveys, administered at the beginning and end of the course, with one following a year later.

After consulting with and obtaining approval from the university’s research ethics board, we announced our intention to conduct the study at the beginning of the semester and put out a call for volunteers. As we quickly realized, although our intention had been to divide the volunteers randomly and evenly between the pilot and traditional sections, we had not enlisted enough volunteers to do so. Although 11 students, all in the traditional lab sections, later volunteered to partake in the anonymized surveys, of the original 20 volunteers, 16 were placed in the pilot section. A further 9 students asked to be placed in the pilot section but declined to participate in the surveys. This meant there were 4 study volunteers out of 88 students in the traditional lab sections (with 11 additional students who completed the surveys), and 16 study volunteers out of 25 students in the pilot lab section.

We realized that this would pose challenges for our quantitative assessment of student outcomes using the new platform and curriculum. Beyond the small sample size and difficulty in obtaining a high student response rate to the surveys, by asking volunteers to actively seek us out at the beginning of the semester to participate, we may have selected for students who were already enthusiastic about electronics or students with above-average confidence in their abilities. This makes it challenging to meaningfully compare the experience of the students in the pilot section with that of their peers in the traditional laboratory sessions. We also note that the pilot section had other differences that could influence student assessments, such as a different teaching assistant delivering the pilot lab who received higher evaluations from students.

Although our ability to make a quantitative assessment of improved learning experiences in the inquiry-based activities is limited, student responses to the \Winterlab platform were positive throughout the laboratory sessions, and this is bolstered by their responses to the set of anonymized surveys. While the students generally felt that the inquiry-based curriculum had been more challenging than that faced by their traditionally-instructed peers, they voiced enthusiasm for the use of the \Winterlab platform, and for being involved in the development of such a project. One described the platform as "[w]ell explained and easy to understand how to use." Others reported enjoying the convenience of the automated functions and integrated data collection, saying that "...[the \Winterlab platform] made the labs a lot easier since we were able to directly export and use data for our work. I enjoyed working with it a lot." Although students did not have the opportunity to participate in both the pilot and traditional-style labs, these comparisons between the two are likely informed by discussions with their peers, and comparisons to previous courses taken.

While, due to the nature of the pilot program, the students were not able to keep the boards permanently, five students borrowed the board over the summer in order to pursue their own creative projects, and one student requested access to the source code for both the board and software interface (which had not previously been made fully available). We regard this outcome -- 6 out of 25 students pursuing independent projects outside the course --- as the most promising outcome of the program.

\section{Reflections}
It is well-established in teaching literature that increasing the independence and autonomy that students encounter in their learning environment builds intrinsic motivation and encourages long-lasting investment in the knowledge acquired. The role of the teaching laboratory, to reinforce concepts seen in class by allowing students to interact with the subject matter in new, dynamic ways, may be augmented by an increased emphasis on independent problem-solving and individual experimentation, both of which are assisted by increased flexibility in when and where learning takes place. The increasing accessibility of low-cost, standalone test and measurement platforms allows students to take their course-based experiments with them and to directly apply the knowledge and skills they acquire to projects and experiments in their lives outside the classroom.

In seeking to apply these principles to an introductory undergraduate signal processing and electronics course with the hope of improving students’ engagement with and enjoyment of the course material, we designed and produced a learning platform that incorporates the functionality of the equipment used in the typical undergraduate electronics laboratory (function generator, oscilloscope, digital multimeter, and voltage supply) within a microcontroller-based unit controlled by a student’s laptop. The \Winterlab platform is portable, robust, costs roughly as much as a textbook to produce, and consists only of circuits and components that are commonly taught at the level of the course for which it was designed.

The \Winterlab board is not the only device of its kind; established electronics manufacturers such as \citet{analogdevices} and \citet{digilent} have released integrated test and measurement platforms with a range of levels of performance and price, and smaller projects such as the Espotek Labrador \cite{esposito} offer friendly open-source avenues. Although these platforms may also be useful to students or hobbyists at this level, the decisions, processes, and motivations contributing to the design of the \Winterlab project increase its relevance as a teaching and learning tool.

Each aspect of the platform was conceived and developed to open doors to extensions beyond its initial use-case. With the software interface, we strove to emulate the look and feel of physical lab equipment, so that students would develop familiarity with tools that are commonly used in experimental science, while still allowing them to automate and streamline their measurements so that they spend more time developing their projects and less time pressing buttons. We chose an affordable, high-performance, Arduino-compatible microcontroller to power the board and integrated it in such a way that the student could easily physically uproot it and repurpose it for other projects. We made sure that the style of its programming was such that it was as easy to read -- and modify – as possible. We worked with students as they used the platform in situ, incorporating their feedback on bugs and features in order to continually improve both the platform’s performance and their experience using it.

While some aspects of the project, particularly the attempt at conducting a quantitative study of the impact of technology-enabled inquiry-based learning in the electronics lab, did not succeed in the manner we had originally envisioned, we learned continually throughout the course of the project. Having set out to encourage students to learn and engage with electronics through independent tinkering, seeing several student participants borrow the platform after the end of the course to pursue their own personal projects was a rewarding indication that the project was successful.

Finally, there is the unique, somewhat introspective nature of this design case: that a platform designed to help students learn through independent exploration would be developed in large part by a fellow student, learning through the exploration of its design. This rich individual learning opportunity, broadened by interactions with students of the course during the project’s development and deployment years, resulted in the production of a versatile and accessible learning platform, well-suited to encourage growth and exploration both within the undergraduate lab and beyond.

\section{Acknowledgements}
The authors acknowledge funding from the Natural Sciences and Engineering Research Council of Canada, the Canadian Institute for Advanced Research, and the Supporting Active Learning \& Technological Innovation in Studies of Education (SATLISE) community.
All procedures performed in this study involving human participants were in accordance with the ethical standards of the institutional research committee, the McGill University Research Ethics Board. Informed consent was obtained from all participants in the study.

\bibliography{bib}

\begin{thebibliography}{15}
\providecommand{\natexlab}[1]{#1}
\providecommand{\url}[1]{\texttt{#1}}
\expandafter\ifx\csname urlstyle\endcsname\relax
  \providecommand{\doi}[1]{doi: #1}\else
  \providecommand{\doi}{doi: \begingroup \urlstyle{rm}\Url}\fi

\bibitem[{Analog Devices}(2021)]{analogdevices}
{Analog Devices}.
\newblock {Adalm1000: Active learning module}, 2021.
\newblock URL
  \url{https://www.analog.com/en/design-center/evaluation-hardware-and-software/evaluation-boards-kits/adalm1000.html}.

\bibitem[Appleton et~al.(2008)Appleton, Christenson, and Furlong]{appleton2008}
J.~Appleton, S.~Christenson, and J.~Furlong.
\newblock Student engagement with school: critical conceptual and
  methodological issues of the construct.
\newblock \emph{Psychology in the Schools}, 45\penalty0 (5):\penalty0 369--386,
  2008.
\newblock \doi{10.1002/pits.20303}.

\bibitem[{Arduino}()]{arduino_uno}
{Arduino}.
\newblock {Arduino Uno Rev3}.
\newblock URL \url{https://store.arduino.cc/products/arduino-uno-rev3/}.

\bibitem[Bouquet et~al.(2017)Bouquet, Bobroff, Fuchs-Gallezot, and
  Maurines]{bouquet2017}
B.~Bouquet, J.~Bobroff, M.~Fuchs-Gallezot, and L.~Maurines.
\newblock Project-based physics labs using low-cost open-source hardware.
\newblock \emph{American Journal of Physics}, 85\penalty0 (3):\penalty0
  216--222, 2017.
\newblock \doi{10.1119/1.4972043}.

\bibitem[Bransford and Schwartz(1999)]{bransford1999}
J.~D. Bransford and D.~L. Schwartz.
\newblock Chapter 3: Rethinking transfer: A simple proposal with multiple
  implications.
\newblock \emph{Review of Research in Education}, 24\penalty0 (1):\penalty0
  61--100, 1999.
\newblock \doi{10.3102/0091732X024001061}.

\bibitem[{Digilent}(2021)]{digilent}
{Digilent}.
\newblock {Analog Discovery 2: 100MS/s USB Oscilloscope, Logic Analyzer and
  Variable Power Supply}, 2021.
\newblock URL
  \url{https://digilent.com/shop/analog-discovery-2-100ms-s-usb-oscilloscope-logic-analyzer-and-variable-power-supply/}.

\bibitem[Esposito()]{esposito}
C.~P. Esposito.
\newblock {EspoTek Labrador Board}.
\newblock URL
  \url{https://espotek.com/labrador/product/espotek-labrador-board/}.

\bibitem[Freeman et~al.(2014)Freeman, Eddy, McDonough, Smith, Okoroafor, and
  Jordt]{freeman2014}
S.~Freeman, S.~L. Eddy, M.~McDonough, M.~K. Smith, N.~Okoroafor, and H.~Jordt.
\newblock Active learning increases student performance in science,
  engineering, and mathematics.
\newblock \emph{Proceedings of the National Academy of Sciences}, 111\penalty0
  (23):\penalty0 8410--8415, 2014.
\newblock \doi{10.1073/pnas.1319030111}.

\bibitem[McDermott and Shaffer(1993)]{mcdermott1993}
L.~C. McDermott and P.~S. Shaffer.
\newblock Research as a guide for curriculum development: An example from
  introductory electricity. part i: Investigation of student understanding.
\newblock \emph{American Journal of Physics}, 60\penalty0 (11):\penalty0
  994--1003, 1993.
\newblock \doi{10.1119/1.17003}.

\bibitem[Michael(2006)]{michael2006}
J.~Michael.
\newblock Where’s the evidence that active learning works?
\newblock \emph{Advances in Physiology Education}, 30\penalty0 (4):\penalty0
  159--167, 2006.
\newblock \doi{10.1152/advan.00053.2006}.

\bibitem[{NXP Semiconductors}(2017)]{nxp}
{NXP Semiconductors}.
\newblock {Kinetis K66 Sub-Family}, 2017.
\newblock URL
  \url{https://www.nxp.com/docs/en/data-sheet/K66P144M180SF5V2.pdf}.

\bibitem[Schultz(2016)]{schultz2016}
K.~D. Schultz.
\newblock Phase-sensitive detection in the undergraduate lab using a low-cost
  microcontroller.
\newblock \emph{American Journal of Physics}, 84\penalty0 (7):\penalty0
  557--561, 2016.
\newblock \doi{10.1119/1.4953341}.

\bibitem[Stoffregen()]{stoffregen}
P.~Stoffregen.
\newblock {Teensy® 3.6 development board. Teensy® USB Development Board}.
\newblock URL \url{https://www.pjrc.com/store/teensy36.html}.

\bibitem[Thornton and Sokoloff(1998)]{thornton1998}
R.~K. Thornton and D.~R. Sokoloff.
\newblock Assessing student learning of newton's laws: The force and motion
  conceptual evaluation and the evaluation of active learning laboratory and
  lecture curricula.
\newblock \emph{American Journal of Physics}, 66\penalty0 (4):\penalty0
  338--352, 1998.
\newblock \doi{10.1119/1.18863}.

\bibitem[Zachariadou et~al.(2012)Zachariadou, Yiasemides, and
  Trougkakos]{zachariadou2012}
K.~Zachariadou, K.~Yiasemides, and N.~Trougkakos.
\newblock A low-cost computer-controller arduino-based educational laboratory
  system for teaching the fundamentals of photovoltaic cells.
\newblock \emph{European Journal of Physics}, 33\penalty0 (6):\penalty0 1599,
  2012.
\newblock \doi{10.1088/0143-0807/33/6/1599}.

\end{thebibliography}

\newpage
\appendix
\section{Functional diagram of the \Winterlab platform}
\begin{figure}
    \centering
    \captionsetup{width=1.1\linewidth}
    \includegraphics[width=\textwidth]{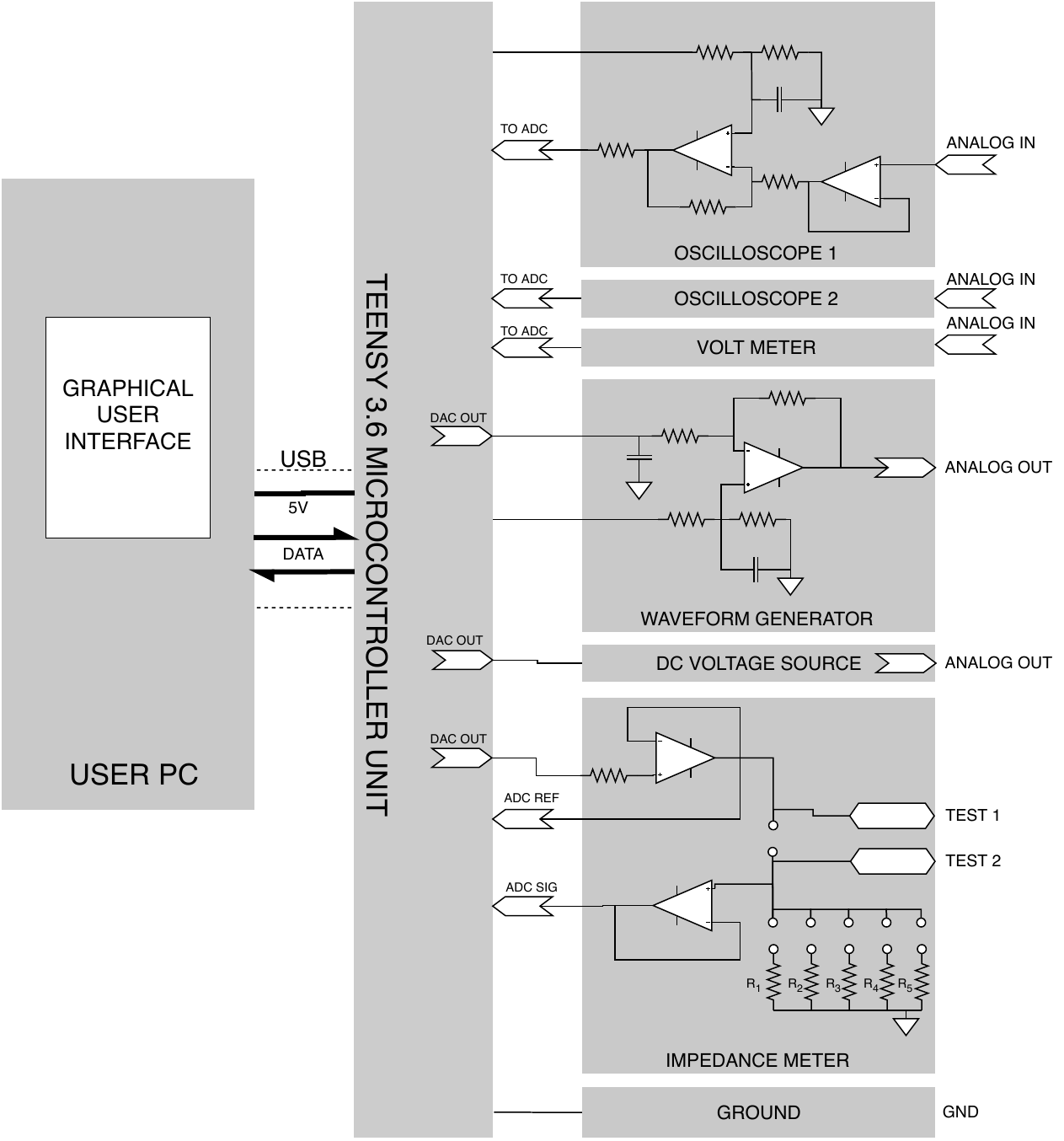}
    \caption{Functional diagram of the \textit{WinterLab} platform, depicting the Teensy$^{\textcircled{R}}$ 3.6 microcontroller unit (top deck of the board; shown at diagram center above) connected to the user’s personal computer (left), and auxiliary circuitry (bottom deck of the board; shown at right) used to enable and protect the various inputs and outputs. The auxiliary circuitry is, from the top of the diagram: two identical oscilloscope channels (inputs), a voltmeter (input; using the same circuit architecture as the oscilloscope channels), a waveform generator (output), a direct current voltage source (output; identical circuit to the waveform generator), and impedance meter (two connection points, across which the user places the component to be measured and selects the impedance measurement range manually amongst R1 through R5). Each section interfaces with the user’s projects via test leads attached to the connection point loops (see Figure \ref{fig:fig6}).}
    \label{fig:appendix}
\end{figure}

\end{document}